\documentclass[11pt]{article}
\usepackage{fullpage}
\usepackage{tikz}
\usepackage{microtype}

\usepackage{ifthen}
\usepackage{graphics,epsfig}
\usepackage{amsmath,amsfonts,amssymb}
\usepackage{algpseudocode}
\usepackage{color}
\usepackage{algorithm}

\usepackage{mathpazo}
\usepackage{textcomp}
\usepackage{nicefrac}
\usepackage{authblk}

\newtheorem{theorem}{Theorem}
\newtheorem{lemma}{Lemma}

\newtheorem{definition}{Definition}
\newcommand{\qed}{\hfill $\Box$ \medbreak}
\newenvironment{proof}{\noindent {\bf Proof.}}{\qed}

\newcommand{\CONGEST}{{\sc congest}}
\newcommand{\cA}{{\mathcal A}}
\newcommand{\cC}{{\mathcal C}}
\newcommand{\cE}{{\mathcal E}}
\newcommand{\cI}{{\mathcal I}}

\newcommand{\cP}{{\mathcal P}}
\newcommand{\cS}{{\mathcal S}}
\newcommand{\cR}{{\mathcal R}}
\newcommand{\cX}{{\mathcal X}}
\newcommand{\ID}{\mbox{\rm ID}}
\newcommand{\accept}{\mbox{\sf accept}}
\newcommand{\reject}{\mbox{\sf reject}}

\begin{document}

\title{Distributed Detection of Cycles}

\author[1]{Pierre Fraigniaud\thanks{Additional support from ANR Project DESCARTES and Inria Project GANG.}}
\author[2]{Dennis Olivetti\thanks{Additional support from ANR Project DESCARTES.}}
\affil[1]{Institut de Recherche en Informatique Fondamentale\\CNRS and University Paris Diderot, France.}
\affil[2]{Gran Sasso Science Institute, L'Aquila, Italy.}
\date{}

\maketitle

\begin{abstract}
Distributed property testing  in networks has been introduced by Brakerski and Patt-Shamir (2011), with the objective of detecting the presence of large dense sub-networks in a distributed manner. Recently, Censor-Hillel et al. (2016) have shown how to detect 3-cycles in a constant number of rounds by a distributed algorithm. In a follow up work, Fraigniaud et al. (2016) have shown how to detect 4-cycles in  a constant number of rounds as well. However, the techniques in these latter works  were shown not to generalize to larger cycles $C_k$ with $k\geq 5$. In this paper, we completely settle the problem of cycle detection, by establishing the following result. For every $k\geq 3$, there exists a distributed property testing algorithm for $C_k$-freeness, performing in a constant number of rounds. All these results hold in the classical \CONGEST\/ model for distributed network computing. Our algorithm is 1-sided error. Its round-complexity is $O(1/\epsilon)$ where $\epsilon\in(0,1)$ is the property testing parameter measuring the gap between legal and illegal instances. 
\end{abstract}


\section{Introduction}	

\subsection{Context}

\subsubsection{Property Testing}

The objective of (sequential) \emph{property testing}~\cite{Goldreich2010} is the design of efficient mechanisms for detecting whether data-struc\-tu\-res satisfy a given property. In the context of networks, a vast literature has been dedicated to testing the presence or absence of specific patterns like triangles, cycles, cliques, etc. (see, e.g., \cite{AlonKKR08,AlonS06,CzumajGRSSS14,GoldreichR02}). A property testing mechanism, a.k.a. \emph{tester}, is a centralized algorithm $\cA$ which is given the ability to probe nodes with queries of the form $\mbox{\sf deg}(i)$ returning the degree of the $i$th node, and $\mbox{\sf adj}(i,j)$ returning the identity of the $j$th neighbor of the $i$th node. Beside its running time, the quality of a tester is typically measured by the number of queries that it must perform before deciding whether or not the network satisfies the considered property.  

Property testing finds its main interest when the problem is relaxed by simply requiring the tester to distinguish between instances satisfying the property, and instances that are \emph{far} from satisfying that property. In the context of networks, several notions of farness have been considered. We consider here the so-called \emph{sparse} model: Given any $\epsilon\in (0,1)$, an $n$-node $m$-edge network $G$ is said to be $\epsilon$-far from satisfying a graph property $\cP$ if adding and/or removing at most $\epsilon m$ edges to/from $G$ cannot result in a network satisfying $\cP$. 

A tester for a graph property $\cP$ is a randomized algorithm~$\cA$ that is required to accept or reject any given network instance, under the following two constraints: 
\begin{itemize}
\item $G$ satisfies $\cP \Longrightarrow \Pr[\cA\;\mbox{accepts}\; G]  \geq  \nicefrac{2}{3}$~;  
\item $G$ is $\epsilon$-far from satisfying $\cP\Longrightarrow\Pr[\cA\;\mbox{rejects}\; G]  \geq  \nicefrac23$.
\end{itemize}
The success guarantee $\nicefrac23$ is arbitrary, as one can boost any success guarantee by repetition. 

In the case of instances which are nearly satisfying $\cP$ but not quite, the algorithm can output either ways.  Hence, a  tester for $\cP$ is a mechanism enabling to detect degraded instances (i.e., instances that are far from satisfying a desired property $\cP$) with arbitrarily large probability, while correct instances  are accepted also with arbitrarily large probability. 

A tester is 1-sided error if 
\begin{itemize}
\item $G$ satisfies $\cP \Longrightarrow \Pr[\cA\;\mbox{accepts}\; G]  = 1$.
\end{itemize}

\subsubsection{Distributed Property Testing}

Distributed property testing has been introduced in~\cite{BrakerskiP11}, and recently revisited in~\cite{Censor-HillelFS16,FraigniaudRST16}. In networks, a \emph{distributed} tester is a distributed algorithm running at every node in parallel (every node executes the same code). After having inspected its surrounding, i.e., the nodes in its vicinity, every node  outputs \accept\/ or \reject. One says that $\cA$ accepts a network $G$ if and only if all nodes output \accept. That is, $\cA$ rejects if at least one of the nodes outputs \reject. 

In this paper, we are focussing on the detection of cycles, one of the most basic and central structures in graph theory, with impact on Ramsey theory and block design. Let $k\geq 3$. A $k$-node cycle, or $k$-cycle for short, is denoted by $C_k$. A network $G$ is $C_k$-free if and only if $G$ does not contain a $k$-node cycle as a subgraph. A case of particular interest is $k=3$, and a $C_3$ is often called triangle. 

It has been shown in~\cite{Censor-HillelFS16} that, in the classical \CONGEST\/ model\footnote{\sl\small The \CONGEST\/ model states that all nodes perform synchronously in a sequence of rounds; At each rounds, messages of $O(\log n)$ bits can be exchanged along the edges of the network.} for distributed computing~\cite{Peleg2010}, there exists a distributed property testing algorithm for triangle-freeness performing in $O(1/\epsilon^2)$ rounds. This result has been extended in~\cite{FraigniaudRST16} where it is proved that there exists a distributed property testing algorithm for $C_4$-freeness performing in $O(1/\epsilon^2)$ rounds as well. 

Perhaps surprisingly, the techniques in~\cite{Censor-HillelFS16,FraigniaudRST16} do not extend to larger cycles. Indeed, using explicit constructions of so-called Behrend graphs, it was proved in~\cite{FraigniaudRST16} that these techniques fail for most values of $k\geq 5$. That is, these techniques cannot result in a tester performing in a constant number of rounds in all graphs, even if the constant is allowed to be a function of $1/\epsilon$. The existence of distributed property testing algorithms for $C_k$-freeness performing in a constant number of rounds was left open for $k$ larger than~4. 

\subsection{Our results}

We completely settle the problem of cycle detection, for every possible length $k\geq 3$. Specifically, we prove that, for every $k\geq 3$, there exists a 1-sided error distributed property testing algorithm for $C_k$-freeness, performing in $O(1/\epsilon)$ rounds in the \CONGEST\/ model. 

Essentially, we reduce the problem of detecting $k$-cycles to the problem of detecting whether a given edge $e$ belongs to some $C_k$. At first glance, the latter problem may seem to be much more simple. Indeed, it does not require to deal with the link congestion caused by the simultaneous testing of several edges. However, the problem remains actually quite challenging as, in the \CONGEST\/ model, even collecting the identities of the nodes at distance~2 from a given node $u$ might be impossible to achieve in $o(n)$ rounds in $n$-node network. Indeed, $u$ may have constant degree, with $\Omega(n)$ neighbors at distance~2. To overcome this difficulty, we proceed by pruning the set of information transmitted between nodes, namely by pruning  the set of candidate cycles passing through the given edge $e$. This pruning is at the risk of discarding candidate cycles that would have turned out to be actual cycles. Nevertheless, our pruning mechanism guarantees  that at least one actual cycle remains in the current set of candidate cycles throughout the execution of the algorithm. 

Interestingly, the use of randomization is limited to the reduction of the general problem of testing $C_k$-freeness to the problem of detecting whether there exists a $k$-cycle passing through a given edge $e$. Indeed, our algorithm solving the latter problem is \emph{deterministic}. In particular, the aforementioned pruning mechanism is deterministic. That is, the existence of an actual cycle passing through $e$ among the restricted set of candidate cycles kept at each round is not a property that holds under some statistical guarantee, but it holds systematically. 

Moreover, our algorithm for testing the existence of a $k$-cycle passing through a given edge $e$ does not rely on the $\epsilon$-farness assumption. That is, even if there is just a single $k$-cycle passing through $e$, that cycle will be detected by our algorithm.  

After the acceptance of the paper, we became aware of the existence of a combinatorial lemma due to Erd\H{o}s et al.~\cite{EHM64}, stating the following. Let $V$ be a set of size $n$, and let us fix two integers $p$ and $q$ with $p+q\leq n$. Then, for any set $F \subseteq {\mathcal P}(V)$ of subsets of size at most $p$ of $V$, there exists a subset $\widehat{F}$ of $F$ of cardinality at most ${p+q \choose p}$ such that, for every set $C \subseteq V$ of size at most $q$, if there is a set $L \in F$ such that $L \cap C = \emptyset$, then there also exists $\widehat{L} \in \widehat{F}$ such that $\widehat{L} \cap C = \emptyset$. This combinatorial result has been used in different contexts, including the design of sequential parametrized algorithms for the longest path problem~\cite{MR808004}. Our technique for detecting cycles can also be viewed as a distributed implementation of this combinatorial lemma.

\subsection{Related Work}

\subsubsection{Property Testing}

The property of $H$-freeness has been the subject of a lot of investigation in classical (i.e., sequential) property testing. 

In the so-called \emph{dense} model, most solutions exploit the \textit{graph removal lemma}, which essentially states that, for every $k$-node graph $H$, and every $\epsilon > 0$, there exists $\delta > 0$ such that every $n$-node graph containing at most $\delta n^k$ (induced) copies of $H$ can be transformed into an (induced) $H$-free graph by deleting at most $\epsilon n^2$ edges. This lemma was first proved for the case $k= 3$, and later generalized to subgraphs $H$ of any size~\cite{ErdosFR86}, and further to induced subgraphs~\cite{AlonFKS00}. It is possible to exploit this lemma for testing the presence of any (induced or not) subgraph of constant size, in constant time. Notice that $\delta$ is a fast growing function of $\epsilon$ and $k$. The growth of the function was later improved in~\cite{AlonS06} under some assumptions. For more details on the graph removal lemma, see~\cite{ConlonF2012}.

Cycle-detection has also been considered in the so-called sparse model. On bounded degree graphs,  cycle-freeness can be tested with  $O(\frac{1}{\epsilon^3} + \frac{d}{\epsilon^2})$ queries~\cite{GoldreichR02} by a $2$-sided error algorithm, where $d$ is the maximum degree of the graph. However, if we restrict ourselves to $1$-sided error algorithms, then the problem becomes harder. A lower bound of $\Omega(\sqrt{n})$ queries was established in~\cite{CzumajGRSSS14}. The same paper presents a  tester requiring $\widetilde{O}(poly(\nicefrac{1}{\epsilon}) \sqrt{n})$ queries in arbitrary graphs, and another tester requiring $\widetilde{O}(poly(\nicefrac{d^k}{\epsilon}) \sqrt{n})$ queries in  graphs with maximum degree~$d$ for detecting cycles of length at least $k$. Detecting triangles requires at least $\Omega(n^{\nicefrac{1}{3}})$ queries, and at most $O(n^{\nicefrac67})$ queries (see \cite{AlonKKR08}). The same lower bound holds for detecting any non bipartite subgraph $H$, and for $2$-sided error algorithms as well. For some specific subgraphs $H$, the lower bound can even be as high as $\Omega(n^{\nicefrac{1}{2}})$.

\subsubsection{Distributed Property Testing}

Distributed property testing has been introduced in~\cite{BrakerskiP11}, and fully formalized in \cite{Censor-HillelFS16}. 

The authors of that latter paper show that, in the dense model, any tester for a \emph{non-disjointed} property can be emulated in the distributed setting with just a quadratic slowdown, i.e., if a sequential tester makes $q$ queries, then it can be converted into a distributed tester that performs in $O(q^2)$ rounds. This simulation exploits the fact that any dense tester can be converted to a tester that first chooses some nodes uniformly at random, gathers their edges, and then performs centralized analysis of the obtained data (see~\cite{GoldreichT03}).

The authors of  \cite{Censor-HillelFS16} also provide distributed testers for the sparse model, showing that it is possible to test triangle-freeness in $O(\nicefrac{1}{\epsilon^2})$ rounds, cycle-freeness in $O(\nicefrac{1}{\epsilon}\log n)$ rounds, and, in bounded degree graphs, bipartiteness in 
\[
O(poly(\nicefrac{1}{\epsilon}\log(\nicefrac{n}{\epsilon})))
\]
rounds. Their work was inspired by \cite{BrakerskiP11}, where a constant-time distributed algorithm for finding a linear-size $\epsilon$-near clique is proposed, under the assumption that the graph contains a linear-size $\epsilon^3$-near clique. (An $\epsilon$-near clique is a set of nodes where all but an $\epsilon$ fraction of pairs of nodes have edges between them).

The result in \cite{Censor-HillelFS16} regarding testing triangle-freeness was extended in \cite{FraigniaudRST16}, where it is shown that, for every 4-node connected graph $H$, there exists a distributed tester for $H$-freeness performing in $O(\nicefrac{1}{\epsilon^2})$ rounds. Also,  \cite{FraigniaudRST16} provides a proof that the approach in \cite{Censor-HillelFS16,FraigniaudRST16} fails to test $C_k$-freeness in a constant number of rounds, whenever $k\geq 5$.

\subsubsection{Distributed Decision}

Distributed property testing fits into the larger framework of \emph{distributed decision}. The seminal paper \cite{NaorS95} was perhaps the first to identify the connection between the ability to locally check the correctness of a solution in a distributed manner, and the ability to design an efficient deterministic distributed algorithm for constructing a correct solution. Since then, there have been a huge amount of contributions aiming at studying variant of distributed decision, in the deterministic setting (see, e.g., \cite{FKP13}),  the anonymous setting (see, e.g., \cite{EPSW14}), the probabilistic setting (see, e.g., \cite{FF15,FGKPP14}), the non-deterministic setting (see, e.g., \cite{GS16,KKP10}), and even beyond (see, e.g., \cite{BAFO17,FFH16}). We refer to \cite{FF16} for a survey on distributed decision. 

\subsubsection{Distributed Cycle Detection}

Cycle detection has been investigated in various parallel and distributed computing frameworks, in particular for its connection to deadlock detection in routing or databases. We refer to, e.g.,  \cite{BT98,Chaudhuri99,Chaudhuri02,OSGH16} for cycle detection in message passing, bulk-synchronization, self-stabilizing, and other models of parallel and distributed computing. 

\section{Model and Definitions}	

\subsection{The {\small CONGEST} Model}

We are considering the classical \CONGEST\/ model for distributed network computing~\cite{Peleg2010}. The network is modeled as a connected simple graph (no self-loops, and no parallel edges). The nodes of the graph are computing entities exchanging messages along the edges of the graph. Nodes are given arbitrary distinct identities (IDs) in a range polynomial in~$n$, in $n$-node networks. Hence, every ID can be stored on $O(\log n)$ bits. 

In the \CONGEST\/  model, all nodes start simultaneously, and execute the same algorithm in parallel. Computation proceeds synchronously, in a sequence of \emph{rounds}. At each round, every node 
\begin{itemize}
\item performs some individual computation,
\item sends messages to neighbors in the network, and 
\item receives messages sent by neighbors. 
\end{itemize}
The main constraint imposed by the \CONGEST\/ model is a restriction of the amount of data that can be transferred between neighboring nodes during a round: messages are bounded to be of $O(\log n)$ bits. 

The $O(\log n)$-bit bound on the message size enables the transmission of a constant number of IDs between nodes at each round. The \CONGEST\/ model is well suited for analyzing the impact of limiting the throughput of a network on its capacity to solve tasks efficiently. The complexity of a distributed algorithm in the \CONGEST\/ model is expressed in number of rounds.

 In this paper, we are mostly interested in solving tasks locally. Hence, we are mainly focussing on the design of algorithms performing in a constant number of rounds in the \CONGEST\/ model. 
 
 \subsection{Distributed Property Testing}
 
\subsubsection{Definition}

Let $\cP$ be a graph property like, e.g., planarity, cycle-freeness, bipartiteness, $C_k$-freeness, etc.  Let $\epsilon\in(0,1)$. Recall that a graph $G$ is said to be $\epsilon$-far from satisfying $\cP$ if removing and/or adding at most $\epsilon m$ edges to/from $G$ cannot result in a graph satisfying $\cP$. 

A distributed property testing algorithm for $\cP$ is a randomized algorithm which performs as follows. Initially, every node is only given its ID as input. After a certain number of rounds, every node must output a value in $\{\accept,\reject\}$. The algorithm is correct if and only if the following two conditions are satisfied. 

\medskip 

\noindent $\bullet$ if $G$ satisfies $\cP$, then 
\[
\Pr[\mbox{every node outputs \accept}]  \geq  \nicefrac{2}{3};
\] 
$\bullet$ if $G$ is $\epsilon$-far from satisfying $\cP$, then 
\[
\Pr[\mbox{at least one node outputs \reject}]  \geq  \nicefrac23.
\]
The algorithm is 1-sided error if, whenever $G$ satisfies $\cP$, the probability that every node outputs \accept\/ equals~1, i.e., if $G$ satisfies $\cP$, then 
\[
\Pr[\mbox{every node outputs \accept}]  = 1.
\]  

\subsubsection{$C_k$-Freeness}

Let $k\geq 3$. A $k$-node cycle, or simply $k$-cycle for short, consists of $k$ nodes $x_i$, and $k$ edges $\{x_i,x_{i+1\bmod k}\}$, $i=0,\dots,k-1$. Such a graph is denoted by~$C_k$. 

Given a graph $G$, its set of nodes (resp., edges) is denoted by $V(G)$ (resp., $E(G)$). Throughout the paper, $n=|V(G)|$, and $m=|E(G)|$. 

Recall that a graph $H$ is a \emph{subgraph} of a graph $G$ if and only~if 
\[
V(H)\subseteq V(G) \; \mbox{and} \; E(H)\subseteq E(G).
\] 

\begin{definition}
A network $G$ is $C_k$-free if and only if $G$ does not contain a $k$-node cycle as a subgraph. 
\end{definition}

Our objective is the design of efficient distributed property testing algorithms for $C_k$-freeness, for all $k\geq 3$. 

\section{Detecting Cycles}	

In this section, we establish our main result. 

\begin{theorem}\label{theo:main}
For every $k\geq 3$, there exists a 1-sided error distributed property testing algorithm for $C_k$-freeness performing in $O(\frac{1}{\epsilon})$ rounds in the \CONGEST\/ model.
\end{theorem}

The rest of the section is dedicated to the proof of the theorem. Let us fix $k\geq 3$. We need to show that there exists a distributed tester for $C_k$-freeness performing in $O(\frac{1}{\epsilon})$ rounds, satisfying

\medskip 

\noindent $\bullet$ if $G$ is $C_k$-free, then 
\[
\Pr[\mbox{every node outputs \accept}]  = 1.
\] 
$\bullet$ if $G$ contains a cycle $C_k$ then 
\[
\Pr[\mbox{at least one node outputs \reject}]  \geq  \nicefrac23.
\] 


\noindent Our tester algorithm for detecting $C_k$ proceeds in two phases: 

\begin{enumerate}
\item determining a candidate edge $e$ susceptible to belong to some cycle $C_k$, if any; 
\item checking the existence of a cycle $C_k$ passing through $e$. 
\end{enumerate}

\noindent Only the first phase is randomized, the second phase is fully deterministic. 

\subsection{Description of Phase~1} 

Every edge is assigned to its extremity with smallest identity. Every node picks a random integer $r(e)\in[1,m^2]$ for each edge $e$ that is assigned to it, called the rank of~$e$. (By construction, $O(\log n)$ random bits per edge are sufficient). For each edge $e$, the extremity of $e$ which computed its rank sends $r(e)$ to the other extremity. Then, every node $u$ selects the edge $e_u$ of lowest rank among all its incident edges, where ties are broken arbitrarily (e.g., based on the ID of extremities), and starts performing the second phase, which consists in checking whether there exists a cycle $C_k$ passing through~$e_u$. 

To avoid congestion, every node performs only instructions of Phase~2 related to the edge with smallest rank it ever become aware of during the execution of the algorithm (again, ties are broken arbitrarily), in a way similar to the prioritized search in~\cite{Censor-HillelFS16}.  Specifically, if a node $u$ currently involved in checking the existence of a cycle $C_k$ passing through $e$ receives a message related to checking the existence of a cycle $C_k$ passing through $e'\neq e$, then $u$ discards this message if 
\[
r(e')>r(e), 
\]
and otherwise switches to checking the existence of cycles passing through $e'$. This guarantees that no two messages corresponding to checking the existence of a cycle $C_k$ passing through two different edges ever traverse an edge in the same direction at the same round. Moreover, if there is a unique edge $e$ with minimum rank, then no nodes discard messages related to checking the existence of a cycle $C_k$ passing through $e$, and thus the checking phase for $e$ will not be interrupted. 

Before analyzing Phase~1, we now describe Phase~2, which is the core of the property testing algorithm for $C_k$-freeness. For simplifying the presentation, let us fix some edge 
\[
e=\{u,v\}, 
\]
and let us describe Phase~2 for edge~$e$ only, assuming that no other checks for other edges are  running concurrently. Likewise, the reader can assume that $e$ is the unique edge with minimum rank in~$G$, which guarantees that the Phase~2 for $e$ will not be slowed down by messages corresponding to Phase~2 applied to other edges.  

\subsection{Description of Phase~2}  

We  describe the algorithm used to check whether there exists a cycle $C_k$ passing through a given edge~$e$.  The algorithm proceeds in $\lfloor \frac{k}{2} \rfloor$ rounds. At each round~$t=1,\dots,\lfloor \frac{k}{2} \rfloor$ of the algorithm, sequences of~$t$ IDs are exchanged between nodes participating to the search for~$C_k$. Every node which receives some sequences at round~$t$ concatenates its own ID to each received sequence, and sends the resulting collection of sequences to all its neighbors. 

\begin{figure}[htb]
\begin{center}
\includegraphics[width=5.5cm]{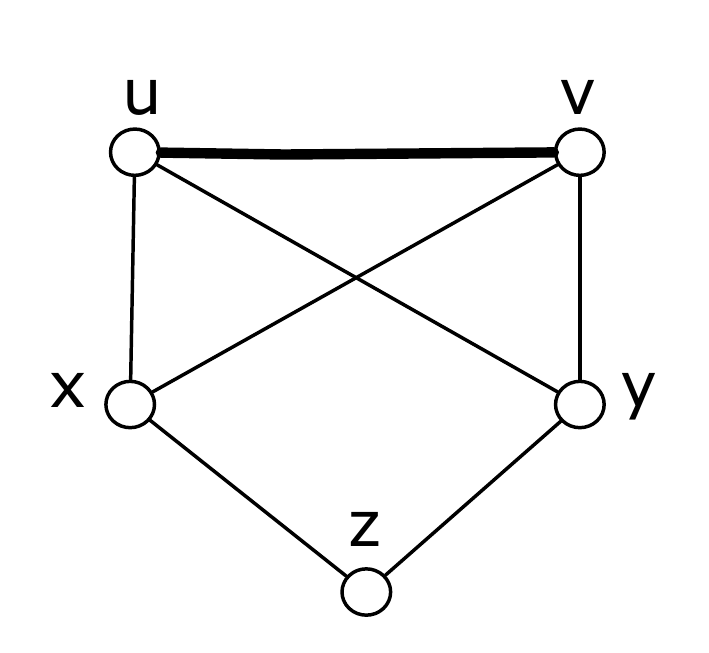}
\caption{\sl Detecting $C_5$ passing through $\{u,v\}$}
\label{fig:C5}
\end{center}
\end{figure}

For instance, to detect a $C_5$ passing through $e=\{u,v\}$, nodes $u$ and $v$ send their IDs to their neighbors at Round~1 (see Fig.~\ref{fig:C5}). A node may thus receive 0, 1, or 2~IDs depending on whether it is adjacent to none, one, or both nodes $u$ and $v$. Each node $x$ which received at least one of these IDs has now a set $\cR$ of sequences of the form $(\ID(w))$ where $w\in \{u,v\}$. Such node appends its own ID to each sequence, and sends the resulting set of sequences to all its neighbors at Round~2. A node $z$ that, at Round~2, receives a sequence $(\ID(u),\ID(x))$, and a sequence $(\ID(v),\ID(y))$ from distinct neighbors $x$ and $y$, respectively, detects the presence of the cycle $(u,x,z,y,v)$. 

This ``append-and-forward'' technique can be trivially extended to detect $C_k$, for arbitrary $k\geq 5$. However, a node of high degree may have to forward very many sequences during a round (this is typically the case of a node connected to $u$ and/or $v$ via many vertex-disjoint paths of same length), violating the bandwidth restriction of the \CONGEST\/ model. 

The main concern of our algorithm is to limit the maximum number of different sequences of IDs to be sent by each node during the execution. Yet, it is crucial that nodes forward sufficiently many sequences of IDs to guarantee detection. For instance, in the graph depicted on Fig.~\ref{fig:C5}, nodes $x$ and $y$ both receive $\ID(u)$ and $\ID(v)$ at the first round. If $x$ forwards only the sequence $(\ID(u),\ID(x))$, and $y$ forwards only the sequence $(\ID(u),\ID(y))$, the 5-cycle will not be detected by~$z$. 

In other words, discarding too many sequences may prevent the algorithm from detecting the cycle, while forwarding too many sequences overloads the communication links. We show that sending a constant number of sequences is sufficient to guarantee cycle detection whenever these sequences are carefully chosen. 

\medskip

The pseudocode of our algorithm is depicted as Algorithm~\ref{ckdet}. 

\subsection{Description of  Algorithm~\ref{ckdet}} 

Algorithm~\ref{ckdet} is essentially of the form ``append-and-forward'' (cf. Instruction~\ref{alg:append}), but selects only a few lists to be sent at each round. The ``seed'' lists are just formed by $\ID(u)$ and $\ID(v)$ (cf. Instruction~\ref{alg:seed}). The algorithm proceeds in $\lfloor \frac{k}{2} \rfloor$ rounds (cf. the for-loop of Instruction~\ref{alg:forloop}). At each round, every node that received non-empty messages collects all IDs that were contained in these messages, distinct from its own ID, in a set~$\cI$ (cf. Instructions~\ref{alg:setoflists}-\ref{alg:setofIDs}). Then, at round~$t$, a set of $k-t$ ``fake'' IDs are added in~$\cI$ (cf. Instruction~\ref{alg:fake}). Intuitively, these  fake IDs represent the yet unknown IDs of nodes which could potentially form a $C_k$ together with the nodes of some list received by the current node at this round. In order to form the collection $\cS$ of lists that will be sent to neighbors at the next round (cf. Instruction~\ref{alg:send}), the collection $\cX$ of all possible sets $X$ of $k-t$ IDs are constructed, including fake IDs (cf. Instruction~\ref{alg:allsetsofIDs}). 

The core of the algorithm is the construction of $\cS$ by the Instructions from~\ref{alg:initR} to~\ref{alg:append}. Before describing this crucial part of  Algorithm~\ref{ckdet} in detail, let us complete the description of the final part of the algorithm.  

At Round~$\lfloor\frac{k}{2} \rfloor$, all lists of IDs sent and received at this round, or received during the previous round, are considered, and stored in a set of lists $\cR$ (cf. Instructions~\ref{alg:consideralllists0}-\ref{alg:consideralllists}). If a node $w$ has two lists $L_1$ and $L_2$ in $\cR$ such that 
\[
|L_1 \cup L_2 \cup \{\ID(w)\}| = k,
\] 
then Node~$w$ outputs ``yes''. Before showing that a cycle $C_k$ formed by all nodes with IDs in $L_1 \cup L_2 \cup \{\ID(w)\}$  exists if and only if $|L_1 \cup L_2 \cup \{\ID(w)\}| = k$, we first return to the core of Algorithm~\ref{ckdet}, that is the set up of the set of lists $\cR$. 

\begin{algorithm*}[]
	\caption{$C_k$ detection for edge $e=\{u,v\}$ executed by node with ID $myid$.}
	\label{ckdet}
	\begin{algorithmic}[1] 
	\Function{detectCk}{$u$,$v$}
		\If{$myid = u$ \textbf{or} $myid = v$} \Comment{\textsf{initial computation at round 1}}
			\State $\cS\gets \{(myid)\}$ \Comment{\textsf{$\cS$ is a set of sequences of IDs}} \label{alg:seed}
		\Else
			\State $\cS\gets\emptyset$
		\EndIf
		\State \textbf{send} $\cS$ \textbf{to} all neighbors \Comment{\textsf{send operation at round $1$}} 
		\State{\textbf{receive} messages \textbf{from} all neighbors} \Comment{\textsf{receive operation at round $1$}}
		
		\medskip
		
		\For{$t=2$ \textbf{to} $\lfloor \frac{k}{2} \rfloor$} \Comment{\textsf{rounds $2$ to $\lfloor \frac{k}{2} \rfloor$}} \label{alg:forloop}
			\If{non-empty messages have been received at round $t-1$}
			\State $\cR \gets$ set of all ordered sequences of IDs received at round $t-1$ \Comment{\textsf{$\cR$ 
				contains sequences of $t-1$ IDs}}\label{alg:setoflists}
 			\State remove from $\cR$ all sequences containing $myid$ \label{alg:myinremoved}
			\State $\cI \gets $ set of IDs included in at least one sequence in $\cR$ \label{alg:setofIDs}
			\State $\cI\gets \cI \cup \{-1,\dots,-k+t\}$ \Comment{\textsf{add $k-t$ distinct ``fake'' IDs to $\cI$}} \label{alg:fake}
			\State $\cX \gets $ collection of all sets $X$ of $k-t$ IDs in $\cI$ \label{alg:allsetsofIDs}
			\State $\cS \gets \emptyset$ \Comment{\textsf{initializes the set of sequences to be sent}} \label{alg:initR}
			\ForAll{ $L \in \cR$ } \label{alg:constructS}
				\State $\cC \gets \{X \in \cX: X \cap L =\emptyset\}$ \Comment{\textsf{$\cC$ is a sub-collection of sets $X$ of $k-t$ IDs}}
				\If{ $\cC\neq \emptyset$ }
					\State $\cS \gets \cS \cup \{L\}$ \Comment{\textsf{$\cS$ contains ordered sequences of existing IDs}}
					\State $\cX \gets \cX \setminus \cC$
				\EndIf
			\EndFor			
			\State append $myid$ at the tail of each $L\in \cS$ \Comment{\textsf{$\cS$ contains sequences of $t$ IDs}}
				\label{alg:append}
		\Else 
			\State $\cS \gets \emptyset$
		\EndIf
		\State \textbf{send} $\cS$ \textbf{to} all neighbors \Comment{\textsf{send operation at round $t$}} \label{alg:send}
		\State{\textbf{receive} messages \textbf{from} all neighbors} \Comment{\textsf{receive operation at round $t$}}
		\EndFor
		
		\medskip
				
		\If{non-empty messages have been received at any round $1,\dots,\lfloor\frac{k}{2} \rfloor$}
				\If{ $k$ is odd } \label{alg:consideralllists0}	
					\State $\cR  \gets  \{\mbox{sequences received at round $\lfloor\frac{k}{2} \rfloor$}\}$ \Comment{\textsf{$\cR$ contains sequences of equal length}}
				\Else
					\State $\cR  \gets \cS \; \cup \; \{\mbox{sequences received at round } \lfloor\frac{k}{2} \rfloor -1\}$ \Comment{\textsf{$\cR$ contains sequences of lengths differing by at most 1}}
				\EndIf \label{alg:consideralllists}		
			\If{ $\exists L_1, L_2 \in \cR : |L_1 \cup L_2 \cup \{myid\}| = k$} \label{alg:yescondition}
				\State output \reject \Comment{\textsf{a $C_k$ has been detected}}
			\Else~output \accept
		\EndIf
		\Else~output \accept
		\EndIf
		\EndFunction
	\end{algorithmic}
\end{algorithm*}

\paragraph{Construction of the set of lists to be sent at each round.} For comfort and ease of reading, we repeat below the instructions performed by Algorithm~\ref{ckdet} for computing the set $\cS$ of ordered sequences to be sent to all neighboring nodes. 

\medskip

\begin{center}
\begin{minipage}{7cm}
$\cS \gets \emptyset$ \\
\textbf{for all} $L \in \cR$ \textbf{do}\\
\hspace*{3ex} $\cC \gets \{X\in \cX:  X\cap L =\emptyset\}$ \\
\hspace*{3ex} \textbf{if} $\cC \neq \emptyset$ \textbf{then} \\
\hspace*{6ex} $\cS \gets \cS \cup \{L\}$ \\
\hspace*{6ex} $\cX \gets \cX \setminus \cC$\\
\hspace*{3ex} \textbf{end if} \\
\textbf{end for} \\
append $myid$ at the tail of each $L\in \cS$ 
\end{minipage}
\end{center}

\medskip

Recall that $\cR$ denotes the set of all ordered sequences $L$ of IDs received at this round, and $\cX$ denotes the collection of all sets of $k-t$ elements in $\cI$, where $\cI$ is the set of all collected IDs at this round, including the fake IDs in $\{-1,-2,\dots,-k+t\}$. 

For each sequence $L\in \cR$ the algorithm takes the decision whether to include $L$ in $\cS$ or not. For this purpose, the algorithm checks whether there is a set $X\subseteq \cI$ with $k-t$ elements which does not intersect $L$. If this is the case, such a list $L$ is added to $\cS$. The intuition is that $L$ (of length $t-1$ at round~$t$) may potentially be extended by adding the current node, plus $k-t$ other nodes, so that to form a cycle $C_k$.  

\begin{figure}[htb]
\begin{center}
\includegraphics[width=7cm]{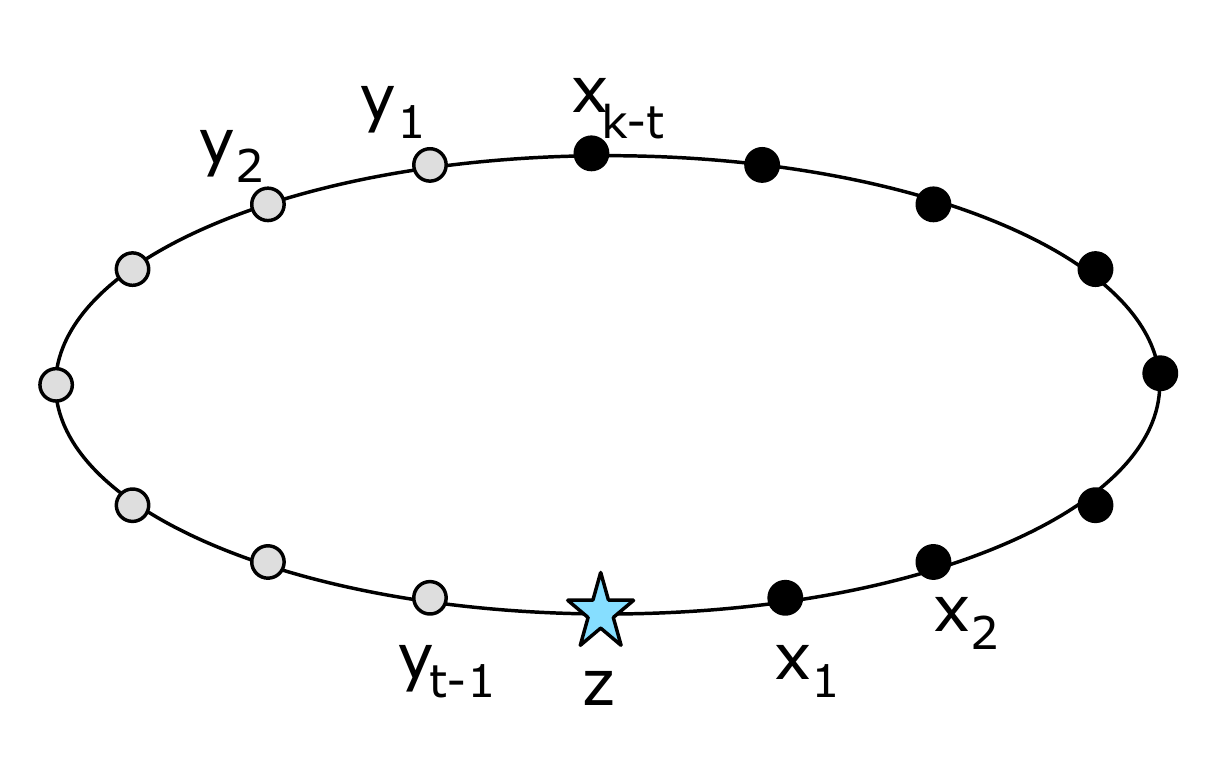}
\caption{\sl Construction of $\cR$}
\label{fig:LcupS}
\end{center}
\end{figure}

For instance, Fig.~\ref{fig:LcupS} displays the case where
\[
L=(y_1,y_2,\dots,y_{t-1})
\]
and 
\[
X=\{x_1,x_2,\dots,x_{k-t}\}
\]
are considered by some node $z$, depicted as a star $\star$ on the figure. The nodes in $L$ are depicted in light grey, while the nodes in $X$ are depicted in black. Note that $X$ is a set, and the ordering of the $x_i$'s on the figure is arbitrary. The list $L$ is placed in $\cS$ because there are $k-t$ nodes, i.e., those in $X$, which can potentially form a $k$-cycle with $z$ and all the nodes in $L$. 

Importantly, the sets $X\in\cC$ are then removed from $\cX$. The intuition is that if there is a $k$-cycle formed by the nodes in $L'\cup\{z\}\cup X$ for some list $L'\in \cR$ where $z$ is the actual node, then  the nodes in $L\cup \{z\} \cup X$ also form a $k$-cycle, and therefore there is no need to keep both $L$ and $L'$. Therefore, as soon as $L$ has been identified, all witnesses sets $X\in \cC$ can safely be removed from $\cX$. 

For instance, considering again the example of  Fig.~\ref{fig:LcupS}, as long as $L$ has been placed in $\cS$, the set $X$ can be removed from $\cX$ since it could only be used to identify another sequence
\[
L'=(y'_1,y'_2,\dots,y'_{t-1})
\]
potentially forming a $k$-cycle with $X$, while we are not interested in enumerating all cycles $C_k$ but just in determining whether there is one. 

Note here the role of the fake IDs that were added to $\cI$. First, observe that the first sequence $L\in\cR$ that is considered in the for-loop (the order in which these sequence are taken is arbitrary) is necessarily placed in $\cS$. Indeed, 
\[
X=\{-1,-2,\dots,-k+t\}
\]
is in $\cX$, and for sure does not intersect $L$. Second, notice that the fact that all sets $X\in \cC$ can be safely removed from $\cX$ is not obvious if $X$ contains fake IDs because $X$ then does not fully specify the cycle. Nevertheless, we shall show that those sets can still be removed, without preventing the algorithm to detect a cycle, if there is one. 

To give a more precise intuition of the use of fake IDs in our algorithm, let us consider a cycle of length $9$, where node IDs are from $1$ to $9$, consecutively around the cycle (hence, the edges are $\{1,2\},\ldots,\{8,9\}$ and $\{1,9\}$. Let us assume that one wants to detect $C_9$, starting from the edge \{1,9\}. Then, in particular, when node $3$ receives the sequence $(1,2)$ from node $2$, we want that node to send the sequence $(1,2,3)$ to node $4$. This is the role of Lines \ref{alg:initR}-\ref{alg:append} in Algorithm \ref{ckdet}, where $\cR$ contains just the sequence $(1,2)$. In Algorithm \ref{ckdet}, if one would not add fake IDs to $\cI$, then $\cI={1,2}$, and $\cX$ would become empty as one cannot construct sequences of length $k-t = 9-3 = 6$ using IDs from $\cI$. As a consequence, $\cC$ would also be empty as it results from an intersection with the empty set, and we would not add $(1,2)$ to $\cS$. It would follow that node $3$ does not send any sequence. Instead, if we add the fake IDs $-1,…,-6$ to $\cI$, then the sequence $(-1,\ldots,-6)$ is in $\cX$, and since $(1,2)$ is disjoint with $\{-1,\ldots,-6\}$, the sequence $(1,2)$ is added to $\cS$, and the sequence $(1,2,3)$ will be sent, as desired. 

\subsection{Analysis of Algorithm~\ref{ckdet}} 

We start by proving the correctness of the algorithm, before analyzing its performances. 

\begin{lemma}\label{lem:seqlength}
For every $t=1,\dots, \lfloor \frac{k}{2} \rfloor$, every sequence $L$ contained in a non-empty set $\cS$ sent at round $t$ is composed of $t$ distinct IDs, and forms a simple path in the graph with one extremity equal to the sender, and the other equal to $u$ or $v$. 
\end{lemma}

\begin{proof}
By induction on $t$. The lemma trivially holds for $t=1$ (cf. Instruction~\ref{alg:seed}). All messages set to be sent at round $t+1$ are constructed by appending the ID of the current node to sequences $L$ received at round~$t$ (cf. Instruction~\ref{alg:append}), and these sequences $L$ do not contain the ID of the current node (cf. Instruction~\ref{alg:myinremoved}). Therefore, every sequence sent at round $t+1$ are composed of $t+1$ distinct IDs. Moreover, by induction, a sequence $L$ received at round $t$ by a node $x$ from a neighboring node $y$ forms a simple path in the graph with one extremity equal to~$y$. Therefore, as long as $\ID(x)\notin L$ (which is guarantied by Instruction~\ref{alg:myinremoved}), the sequence $L\cup\{\ID(x)\}$ forms a simple path in the graph with one extremity equal to~$x$. The other extremity remains unchanged, and thus equal to $u$ or $v$. 
\end{proof}


\begin{lemma}\label{lem:correctness}
For any graph $G$, and every edge $e=\{u,v\}$ of~$G$, Algorithm~\ref{ckdet} running on $G$ satisfies that all nodes output \accept\/ if and only if there are no $C_k$ passing through the edge~$e$. 
\end{lemma}

\begin{proof}
Let us assume that some node $w$ outputs \reject, and let us show that there is indeed a $k$-cycle passing through~$e$. From Instruction~\ref{alg:yescondition}, this node $w$ satisfies that there exist two sequences $L_1, L_2 \in \cR$ such that 
\[
|L_1 \cup L_2 \cup \{\ID(w)\}| = k.
\] 
By Lemma~\ref{lem:seqlength}, both sequences are simple paths of length at most~$\lfloor \frac{k}{2} \rfloor$ from $u$ or $v$ to a neighbor of $w$. Let 
\[
L_1 = (x_1,x_2,\ldots,x_\ell),
\]
and 
\[
L_2 = (y_1,y_2,\ldots,y_m),
\]
where $\ell \leq \lfloor k/2 \rfloor$ and $m\leq \lfloor k/2 \rfloor$. 

\begin{itemize}
\item If $k$ is odd, $|L_1 \cup L_2 \cup \{\ID(w)\}| = k$ implies that $\ell=m= \lfloor k/2 \rfloor$, $w$ is distinct from every $x_i$ and every $y_j$, and  every $x_i$ is distinct from every $y_j$, $i=1,\dots,\ell$, $j=1,\dots,m$. In particular, since $x_1\neq y_1$, we have $\{x_1,y_1\}=\{u,v\}$. It follows that 
\[
(x_1,x_2,\ldots,x_\ell,w,y_m,y_{m-1},\ldots,y_1)
\]
is a $k$-cycle passing through $e$.

\item If $k$ is even, then we claim that 
\[
L_1 \in \cS \; \mbox{and} \; L_2 \not\in \cS
\]
or  
\[
L_1 \not\in \cS \; \mbox{and} \;  L_2 \in \cS.
\] 
Indeed, let us consider two distinct sequences $L$ and $L'$ in $\cS$. Since they are both of length $k/2$,  and since they both contain $\ID(w)$, we have  $|L \cup L' \cup \{\ID(w)\}|\leq k -1$. Thus, at least a sequence must not be contained in $\cS$. Moreover, if  $|L \cup L' \cup \{myid\}| = k$, then at least one of the two sequences must belong to $\cS$ because the sequences received at round $k/2 -1$ are of length $k/2 -1$. Hence the claim holds. 

So, let us now assume, w.l.o.g., that $L_1 \in \cS$ and $L_2 \not\in \cS$. It follows that $L_1$ is of length $k/2$ and contains $\ID(w)$, and that $L_2$ is of length $k/2$ without containing $\ID(w)$. The equality $|L_1 \cup L_2 \cup \{\ID(w)\}| = k$ then implies that $w$ is distinct from every $x_i$ and every $y_j$, and every $x_i$ is distinct from every $y_j$, $i=1,\dots,\ell$, $j=1,\dots,m$. In particular, since $x_1\neq y_1$, we have $\{x_1,y_1\}=\{u,v\}$. It follows that 
\[
(x_1,x_2,\ldots,x_\ell,w,y_m,y_{m-1},\ldots,y_1)
\]
is a $k$-cycle passing through $e$.
\end{itemize}

Therefore, for both cases, $k$ even or odd, the existence of a node which outputs \reject\/ implies the existence of a cycle passing through $e$. 

\bigskip

Conversely, let us assume that there is a $k$-cycle passing through~$e$, and let us show that at least one node detects that cycle (i.e., outputs \reject). Observe that a modified version of the algorithm where the construction of $\cS$ in the for-loop of Instruction~\ref{alg:constructS} is replaced by 
\[
\cS\leftarrow \cR
\] 
clearly detects the cycle. Indeed, at each round $t$, all the possible paths of length $t$ from the edge to the actual node are transmitted. However, there can be too many such paths, and transmitting all of them would not fit with the constraints of the \CONGEST\/ model. Hence, some paths are discarded by Algorithm~\ref{ckdet}. Yet, we show that Algorithm~\ref{ckdet} keeps sufficiently many options for detecting the cycle. Let us fix some round $t\in\{2,\dots,\lfloor\frac{k}{2}\rfloor\}$, and a node $w$. Let us consider a discarded sequence
\[
L=(x_1,x_2,\dots,x_{t-1})
\]
 at $w$. Let us assume that the cycle includes that sequence of nodes, that is the cycle is of the form 
\[
x_1,x_2,\ldots,x_{t-1},w,y_1,\ldots,y_{k-t}
\]
where $\{x_1,y_{k-t}\}=\{u,v\}$. Since the sequence has been discarded, we have 
\[
\{X \in \cX: X \cap L =\emptyset\}=\emptyset
\]
where $\cX$ is the collection of all sets $X$ of $k-t$ IDs in $\cI$, and $\cI$ is the collection of all IDs included in at least one sequence in $\cR$, complemented with the $k-t$ ``fake'' IDs 
\[
\{-1,\dots,-k+t\}.
\] 
This implies that all sets $X\in\cX$ that intersect $L$ have been removed from $\cX$ when considering other sequences in the for-loop. In particular the set 
\[
X=\{y_1,\ldots,y_{k-t}\}
\] 
has been removed when considering another sequence 
\[
L'=(z_1,z_2,\ldots,z_{t-1}). 
\]
Since $L'\cap \{y_1,\ldots,y_{k-t}\}=\emptyset$, we get that there is actually  another cycle, 
\[
z_1,z_2,\ldots,z_{t-1},w,y_1,\ldots,y_{k-t}
\]
where $\{z_1,y_{k-t}\}=\{u,v\}$. Therefore, Algorithm~\ref{ckdet} satisfies that, at every round $t\in\{2,\dots,\lfloor\frac{k}{2}\rfloor\}$, if $w$ belongs to a cycle 
\[
x_1,x_2,\ldots,x_{t-1},w,y_1,\ldots,y_{k-t}
\]
passing through $e=\{x_1,y_{k-t}\}$, and $w$ receives the sequence $x_1,x_2,\ldots,x_{t-1}$, then Algorithm~\ref{ckdet} guarantees that if $w$ does not send the sequence $x_1,x_2,\ldots,x_{t-1},w$ to $y_1$, then $w$ necessarily sends another sequence $z_1,z_2,\ldots,z_{t-1},w$ to $y_1$ where 
\[
z_1,z_2,\ldots,z_{t-1},w,y_1,\ldots,y_{k-t}
\]
is a cycle passing through $e=\{z_1,y_{k-t}\}$. Therefore, the nodes antipodal to $e$ (that is, the nodes at distance $\lceil\frac{k}{2}\rceil-1$ from $e$ in the cycle) will detect a cycle at round $\lfloor\frac{k}{2}\rfloor$, and will output \reject, as desired. 
\end{proof}

In the next lemma, we show that, for a fixed $k$, the messages exchanged during the execution of Algorithm~\ref{ckdet}  are of constant size. 

\begin{lemma}\label{lem:boundedmessage}
For every $t=1,\dots, \lfloor \frac{k}{2} \rfloor$, every message sent by nodes at round~$t$ is composed of at most $(k-t+1)^{t-1}$ ordered sequences of $t$ IDs.
\end{lemma}

\begin{proof}
For the ease of notation, we rephrase the statement of the lemma as: For every $t=0,\dots, \lfloor \frac{k}{2} \rfloor-1$, every message sent by nodes at round~$t+1$ is composed of at most $(k-t)^t$ ordered sequences of $t+1$ IDs. 

Let us fix $t\in\{0,\dots,\lfloor \frac{k}{2} \rfloor-1\}$, and a node $w$, and let us focus on round $t+1$. For $i=0,\dots,t$, let us then define property $P_i$ stating:
\begin{center}
 for every set of $t-i$ IDs,   $w$ sends at most \\ $(k-t)^i$ sequences that contain that set.  
 \end{center}
 
 Note that Property $P_t$ establishes the lemma. 
 
 \medskip

Property $P_0$ stating that, for every set of $t$ IDs, $w$ sends at most one sequence that contains that set, follows from the fact that, in the construction of $\cS$ in the for-loop of Instruction~\ref{alg:constructS}, this set will be sent only once, in one of all its possible orderings. 

\medskip

Let us assume that $P_{i-1}$ holds, and let us establish $P_i$. Consider the case where, during the execution of the for-loop of Instruction~\ref{alg:constructS},  we already added $(k-t)^i$ sequences to $\cS$ containing the same $t-i$ elements $x_1, x_2, \ldots, x_{t-i}$. That is, $\cS$ contains 
\begin{align*}
& \{x_1, x_2, \ldots, x_{t-i},y_{1,1},y_{1,2},\ldots,y_{1,i}\} \\
& \{x_1, x_2, \ldots, x_{t-i},y_{2,1},y_{2,2},\ldots,y_{2,i}\} \\
& \hspace{2cm} \vdots \\ 
& \{x_1, x_2, \ldots, x_{t-i},y_{(k-t)^i,1},y_{(k-t)^i,2},\ldots,y_{(k-t)^i,i}\} 
\end{align*}
After these sequences have been added to $\cS$, the remaining sequences in $\cX$ must contain at least one element of each such sequences. That is, for every $X \in \cX$, 
\[
(x_1 \in X) \lor (x_2 \in X) \lor \ldots \lor (x_{t-i} \in X)
\]
or 
\begin{equation}\label{eq:theys}
\bigwedge_{j=1}^{(k-t)^i}  \Big ( (y_{j,1} \in X) \lor (y_{j,2}\in X)  \lor \ldots \lor (y_{j,i} \in X) \Big ) 
\end{equation}
Indeed, if a sequence does not contain $x_1, x_2, \ldots, x_{t-i}$, then it should contain an element $y_{a,b}$ for each sequence. We can now apply the induction hypothesis to show that the same element $y_{a,b}$ cannot appear more than $(k-t)^{i-1}$ times. Indeed, the sequence $x_1, x_2, \ldots, x_{t-i}, y_{a,b}$ is of length $t-(i-1)$, and therefore, by induction, it cannot appear more than $(k-t)^{i-1}$ times. 

Therefore, since there are $(k-t)^i=(k-t)^{i-1} \cdot (k-t)$ sequences in $\cS$ containing the same $t-i$ elements $x_1, x_2, \ldots, x_{t-i}$, Eq.~\eqref{eq:theys} implies that  a sequence $X \in \cX$ must contain $k-t$ different elements. However, sequences in $X$ are of size $k-t-1$. Therefore, the formula in Eq.~\eqref{eq:theys} cannot be satisfied. It follows that, for every $X \in \cX$, 
\[
(x_1 \in X) \lor (x_2 \in X) \lor \ldots \lor (x_{t-i} \in X).
\]
Let us now consider another sequence 
\[
L=(x_1, x_2, \ldots, x_{t-i},z_1,\ldots,z_i)
\]
taken from $\cR$. This sequence will not be added to $\cS$ because every sequence $X\in \cX$ contains at least one element from $\{x_1, x_2, \ldots, x_{t-i}\}$, which implies that $L\cap X\neq \emptyset$.
 \end{proof}

\subsection{Proof of Theorem~\ref{theo:main}} 

Let us first compute the probability of detecting a cycle in a network which is $\epsilon$-far from being $C_k$-free. We exploit the fact that, in such a network, there must be many edge-disjoint copies of $C_k$, as stated below: 

\begin{lemma}[\cite{FraigniaudRST16}] \label{disjointcopies}
Let $H$ be any graph. Let $G$ be an $m$-edge graph that is $\epsilon$-far from being $H$-free. Then $G$ contains at least $\epsilon m / |E(H)|$ edge-disjoint copies of $H$.
\end{lemma}

Hence, a graph $G$ that is $\epsilon$-far from being $C_k$-free contains at least $\epsilon m/k$ edge-disjoint copies of $C_k$, i.e., $\epsilon m$ edges belong to edge-disjoint cycles. 

\begin{lemma}
The probability that there is a unique edge with minimum rank after the execution of Phase~1 is at least $1/e^2$. 
\end{lemma}

\begin{proof}
The probability that there are no collisions while choosing for each edge a random number from $[1,m^2]$ is 
\begin{eqnarray*}
\frac{m^2-1}{m^2} \times \ldots \times \frac{m^2-m}{m^2} & \ge & \left (\frac{m^2-m}{m^2} \right)^m \\
  & = & \left (1-\frac{1}{m} \right)^m \\
  & \ge & \left (e^{\frac{-2}{m}} \right)^m \\
  & = & \frac{1}{e^2}
\end{eqnarray*}
 where the last inequality holds whenever $m\geq 2$. 
\end{proof}

Let $G$  be a graph that is $\epsilon$-far from being $C_k$-free. Let $\cE$ be the event: 
\begin{center}
there is a unique edge with minimum rank after the \\ execution of Phase~1, and this edge belongs to a $k$-cycle. 
\end{center}
Combining the previous two lemmas, we get that 
\[
\Pr[\cE] \geq \epsilon/e^2.
\]
Now, if event $\cE$ holds, then, by Lemma~\ref{lem:correctness}, at least one node  will output \reject, as desired. To boost the probability of detecting a cycle in a graph that is $\epsilon$-far from being $C_k$-free, we repeat the whole process $\frac{e^2}{\epsilon}\ln 3$ times. In this way, the probability that $\cE$ holds in at least one of these repetitions is at least $\nicefrac23$ as desired. 

By Lemma~\ref{lem:boundedmessage}, each repetition of the whole process of executing Phases~1 and~2 requires a constant number of rounds. This completes the proof of Theorem~\ref{theo:main}. \qed

\section{Conclusion}	

In this paper, we have proved that, for every $k\geq 3$, there exists a 1-sided error distributed property testing algorithm for $C_k$-freeness, performing in $O(1/\epsilon)$ rounds. We mention hereafter some possible directions for further work. 

It was proved in~\cite{FraigniaudRST16} that, for every graph pattern $H$ with at most~4 nodes, there exists a distributed property testing algorithm for $H$-freeness, performing in constant number of rounds, and the question of whether a distributed property testing algorithm for $H$-freeness exists for every arbitrarily large pattern $H$ was left open in~\cite{FraigniaudRST16}. The techniques in this paper does not seem to extend to arbitrary patterns. To see why, consider $H$ as a $k$-cycle with a chord between two nodes.  The pruning technique in Algorithm~\ref{ckdet} discarding some sequences of nodes is oblivious to the neighborhood of the nodes in these sequences. Hence, while  Algorithm~\ref{ckdet} makes sure to keep at least one sequence corresponding to a cycle, if such cycle exists, it may well discard the sequence corresponding to the cycle in $H$, and keep a sequence without a chord. It was also pointed out in~\cite{FraigniaudRST16} that their techniques do not seem to extend to \emph{induced} subgraphs\footnote{\sl\small A graph $H$ is an \emph{induced} subgraph of a graph $G$ iff $V(H)\subseteq V(G)$ and $E(H)=E(G[V(H)])$, i.e., for every $(u,v)\in V(H)\times V(H)$, we have $\{u,v\}\in E(H) \iff \{u,v\}\in E(G)$. (In other words, $H$ is isomorphic to the subgraph of $G$ induced by the nodes in $H$).}. The same apparently holds for the techniques in this paper. The reasons are the same as for detecting a given graph pattern $H$. Indeed, our pruning mechanism is not adapted to detect an induced cycle. It may  well discard a sequence corresponding to the induced cycle, and keep a sequence with chords.

We believe that proving or disproving the existence of distributed property testing algorithms for $H$-freeness, as a subgraph or as an induced subgraph, are potentially challenging but definitely rewarding issues whose study is susceptible to shed new light on the \CONGEST\/ model, and, more generally, to improve our understanding of local distributed computing in presence of bandwidth limitation. 

\bibliographystyle{plain}
\bibliography{biblio}

\end{document}